**UNIVERSITATEA "DUNĂREA DE JOS" GALAȚI**
Școala doctorală a Facultății de Știința Calculatoarelor
Domeniul de doctorat: Ingineria Sistemelor

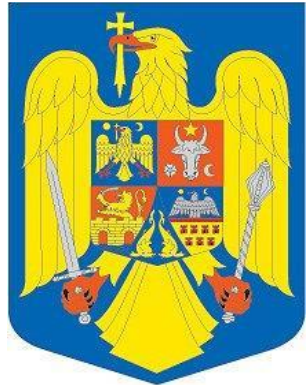

# Proiect de cercetare privind creșterea securității sistemelor informatice și de comunicații prin criptografia cuantică

Conducător,
Prof.dr.ing. Adrian Filipescu

Doctorand,
ing. Cătălin Anghel

- 2009 -

# *Cuprins*

# Abstract

Principalul punct slab al unui sistem criptografic și de comunicații este acela că orice transmisie securizată poate fi făcută numai după ce cheia de criptare este comunicată în secret printr-un canal de comunicații securizat. Cu alte cuvinte, avem de a face cu un paradox : „ Înainte de a comunica în secret, trebuie să comunicăm în secret". Aici intervine criptografia cuantică care profitând de anumite fenomene ce au loc la nivel subatomic nu numai că face imposibilă interceptarea transmisiei dar poate și detecta dacă un atacator ascultă canalul de comunicații. Astfel, având la bază primul protocol al criptografiei cuantice, descoperit de Bennet și Brassard în 1984 [1] (protocolul BB84), propun implementarea unui algoritm criptografic cuantic, care se dorește a fi un protocol de rețea cuantic, pentru a realiza o legătură cuantică între două sisteme informatice și de comunicații.

## Introducere

Informația a însemnat întotdeauna putere, prin urmare dorința de a o proteja, de a o face accesibilă doar unor elite, unor inițiați, s-a pus din cele mai vechi timpuri. Primele texte cifrate descoperite până în prezent datează de circa 4000 de ani și provin din Egiptul Antic.

Există date privind utilizarea scrierii cifrate în Grecia Antică încă din secolul al V-lea î.e.n. Pentru cifrare se folosea un baston în jurul căruia se înfășura, spirală lângă spirală, o panglică îngustă de piele, papirus sau pergament, pe care, paralel cu axa, se scriau literele mesajului. După scriere, panglica era derulată, mesajul devenind indescifrabil. El putea fi reconstituit numai de către persoana care avea un baston identic cu cel utilizat la cifrare. În Roma Antică, secretizarea informațiilor politice și militare se făcea utilizând diverse tipuri de scrieri secrete; amintim cifrul lui Cesar, utilizat încă din timpul războiului galic.

Contribuția arabă la dezvoltarea criptologiei, mai puțin cunoscută și mediatizată, este de o remarcabilă importanță. David Kahn, unul dintre cei mai de seamă istoriografi ai domeniului, subliniază în cartea sa *The Codebreakers* că, criptologia s-a născut în lumea arabă. Primele trei secole ale civilizației islamice (700-1000 e.n.) au constituit, pe lângă o mare extindere politică și militară și o epocă de intense traduceri în limba arabă ale principalelor opere ale antichității grecești, romane, indiene, armene, ebraice și siriene. Unele cărți sursă erau scrise în limbi deja moarte, deci reprezentau în fapt texte cifrate, astfel încât traducerea lor constituie primii pași în criptanaliză, deci originile criptologiei pot fi atribuite arabilor. Dezvoltările criptanalizei au fost mult sprijinite de studiile lingvistice ale limbii arabe. Arabii au preluat cunoștințele matematice ale civilizațiilor grecești și indiene. Arabii sunt cei care au introdus sistemul zecimal de numerotație și cifrele "arabe". Termenii "zero ", "algoritm", "algebră" li se datorază tot lor. Însuși termenul de "cifru" ne vine de la arabi. El provine de la cuvântul arab "sifr" care reprezintă traducerea în arabă a cifrei zero din sanscrită. Conceptul de „zero" a fost deosebit de ambiguu la începuturile introducerii lui în Europa, în care sistemul de numerotație folosit era cel roman. De aceea se obișnuia să se spună despre cineva care vorbea neclar că vorbeste ambigu, ca un cifru. Acest înțeles de ambiguitate a unui mesaj poartă și azi denumirea de cifru. Prin urmare, putem concluziona că, încă din antichitate s-a încercat securizarea informației și a datelor transmise.

## Securitatea sistemelor informatice

Apariția și dezvoltarea continuă a utilizării calculatoarelor în toate domeniile vieții, existența și evoluția rețelelor informatice de comunicații la nivel național și internațional, globalizarea comunicațiilor, existența unor baze de date puternice, apariția și dezvoltarea comerțului electronic, a poștei electronice, pe scurt, dezvoltarea internetului, constituie premisele societății informaționale în care trăim. Toate acestea indică o creștere extraordinară a volumului și importanței datelor transmise sau stocate și implicit a vulnerabilităților acestora. Protecția acestor sisteme, de transmitere și stocare a datelor, presupune existența unor servicii de rețea care să asigure securitatea datelor, cum ar fi :

- *Confidențialitatea*
- *Autenticitatea*
- *Integritatea*
- *Nerepudierea*
- *Controlul accesului*
- *Disponibilitatea*

◆ *Confidențialitatea* este serviciul care are rolul de a proteja datele de atacurile pasive, adică de interceptarea datelor de persoane neautorizate. Se pot identifica mai multe nivele de protecție a acestui serviciu. Cel mai larg nivel a acestuia protejează datele transmise de toți utilizatorii unui sistem. Se pot defini și nivele mai restrânse care protejează doar un utilizator, un mesaj sau chiar unele porțiuni dintr-un mesaj. Acestea pot fi mai puțin folositoare decât abordarea pe scară largă și sunt deseori mai complexe și mai costisitor de implementat. A doua problemă legată de confidențialitate este protecția fluxului informațional de „analiza" traficului. Acest lucru implică ca atacatorul să nu poată identifica sursa, destinația, frecvența, lungimea și alte caracteristici ale datelor transmise prin rețea.

◆ *Autenticitatea* este serviciul legat de garantarea autenticității comunicației. În cazul unui singur mesaj, cum ar fi un semnal de avertisment, funcția serviciului de autenticitate este de a garanta destinatarului că sursa mesajului este aceea care se pretinde a fi. În cazul unei interacțiuni, cum ar fi conectarea unui terminal la un server, două aspecte sunt implicate. În primul rând la inițierea conectării serviciul trebuie să se asigure că cele două părți sunt autentice, adică sunt ceea ce pretind a fi. În al doilea rând, serviciul trebuie să se asigure că comunicația nu interferează cu o a treia parte care s-ar putea preface ca fiind una din cele două părți participante la transmisie, obținând informații nelegitime.

◆ *Integritatea* este serviciul care trebuie să asigure recepționarea mesajelor așa cum au fost transmise fără copierea, inserția, modificarea, rearanjarea sau retransmiterea acestora. Acest serviciu include și protecția împotriva distrugerii datelor. El mai poate include și recuperarea datelor după un atac.

◆ *Nerepudierea* împiedică atât expeditorul cât și destinatarul de a nega transmiterea sau recepționarea unui mesaj. Când un mesaj este trimis, destinatarul poate dovedi că mesajul a fost trimis de pretinsul expeditor. Similar, când un mesaj este recepționat, expeditorul poate dovedi că mesajul a fost recepționat de realul destinatar.

◆ *Controlul accesului* este abilitatea de a limita și controla accesul la sisteme gazdă și aplicații prin legături de comunicație. Pentru a realiza acest control, fiecare entitate care încearcă să obțină acces la un sistem trebuie mai întâi identificată și autentificată după care i se aplică drepturile de acces individuale.

◆ *Disponibilitatea* se referă la asigurarea că sistemele de calcul sunt accesibile utilizatorilor autorizați când și unde acestia au nevoie și în forma necesară, adică informația stocată electronic este unde trebuie să fie, când trebuie să fie și în forma în care trebuie să fie.

### Atacuri asupra securitații sistemelor informatice

Informația care circulă într-un sistem de transmitere și stocare a datelor are un flux normal adică de la sursă la destinație. Atacurile asupra securității sistemelor de transmitere și stocare a datelor sunt acele acțiuni care interceptează, modifică, distrug sau întârzie fluxul normal de date. O clasificare a acestor atacuri este reprezentată în figura 1 :

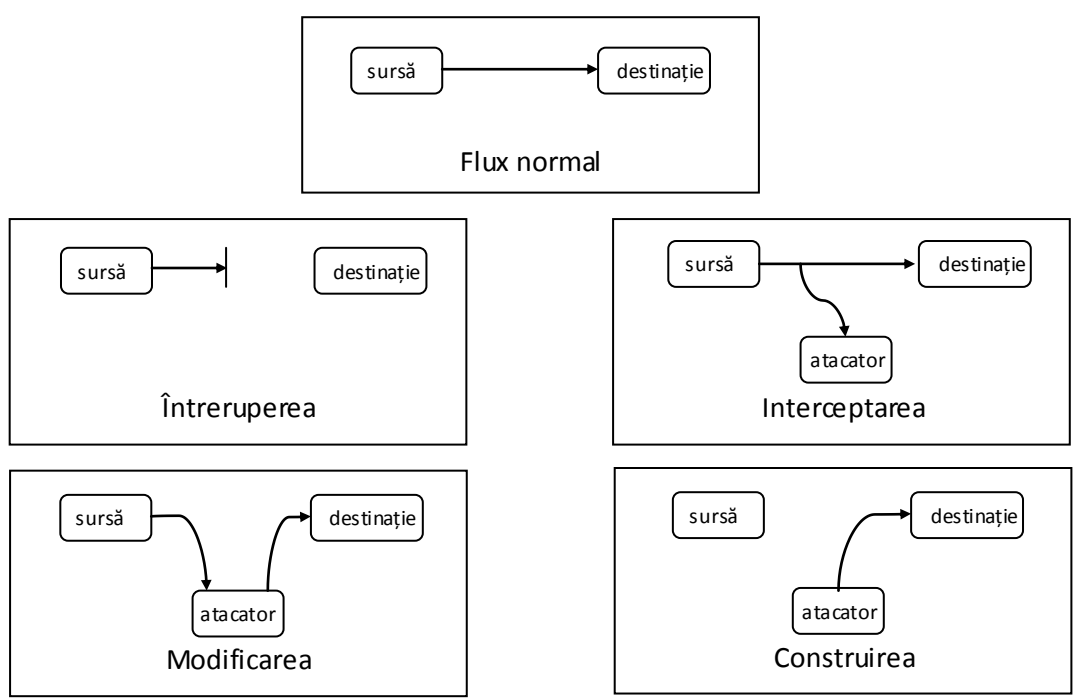

Figura 1. Clasificarea atacurior

- *Întreruperea:* O componentă a sistemului este distrusă, devine indisponibilă sau inutilizabilă total sau pentru o anumită perioadă de timp. Acest tip de atac este un atac asupra disponibilității. Exemple de astfel de atacuri ar fi: distrugerea unor echipamente hardware, tăierea liniilor de comunicație, distrugerea sistemului de fișiere sau tehnici de supraîncărcare a sistemului, care duc la blocarea comunicațiilor sau a întregului sistem, numite atacuri prin refuzul serviciilor (denial of service).
- *Interceptarea:* Inamicul obține acces la o componentă a sistemului. Acesta este un atac asupra confidențialității. Inamicul poate fi o persoană, un program sau un calculator. Exemple de astfel de atacuri ar fi copierea ilegală a unor fișiere și programe sau interceptarea liniilor de comunicație.

◆ *Modificarea:* Inamicul obține nu numai acces la o componentă din sistem, dar și falsifică informația obținută. Acesta este un atac asupra integrității. Exemplele includ modificarea unor valori din fișiere de date, modificarea unor programe sau transmiterea unor mesaje false prin rețea.

◆ *Construirea:* Inamicul pătrunde în sistem și imită unele componente din acesta. Un atac de acest tip este un atac asupra autenticității. Un exemplu de acest gen ar putea fi introducerea unor mesaje false în rețea care sunt interpretate ca mesaje reale și inventarea unor fișiere care pot induce în eroare utilizatorii reali.

O altă clasificare a atacurilor asupra securității sistemelor de transmitere și stocare a datelor poate împărți acestea în două categorii : atacuri pasive și atacuri active.

◆ *Atacuri pasive:* sunt acelea în cadrul cărora inamicul *interceptează* informația ce trece prin canalul de comunicație, fără să interfereze cu fluxul sau conținutul mesajelor. Ca urmare, se face doar analiza traficului, prin citirea identității părților care comunică și învățând lungimea și frecvența mesajelor vehiculate; chiar dacă conținutul acestora este neinteligibil, poate iniția ulterior alte tipuri de atacuri. Atacurile pasive au urmatoarele caracteristici comune:

> ➢ Nu cauzează pagube (nu se șterg sau se modifică date);
> ➢ Încalcă regulile de confidențialitate;
> ➢ Obiectivul este de a "asculta" datele care circulă prin rețea;
> ➢ Pot fi realizate printr-o varietate de metode, cum ar fi supravegherea legăturilor telefonice sau radio, exploatarea radiațiilor electromagnetice emise, rutarea datelor prin noduri adiționale mai puțin protejate.

◆ *Atacuri active*: sunt acelea în care inamicul se angajează fie în furtul mesajelor, fie în modificarea, reluarea sau inserarea de mesaje false. Aceasta înseamna ca el poate șterge, întârzia sau modifica mesaje, poate să insereze mesaje false sau vechi, poate schimba ordinea mesajelor, fie pe o anumită direcție, fie pe ambele direcții ale unui canal de comunicații. Aceste atacuri sunt serioase deoarece modifică starea sistemelor de calcul, a datelor sau a sistemelor de comunicații.

Există următoarele tipuri de amenințări active:

- *Mascarada* - este un tip de atac în care o entitate pretinde a fi o altă entitate. De exemplu, un utilizator încearcă să se substituie altuia sau un serviciu pretinde a fi un alt serviciu, în intenția de a lua date secrete (numărul cărții de credit, parola sau cheia algoritmului de criptare). O mascarada este însoțită, de regulă, de o altă amenințare activă, cum ar fi înlocuirea sau modificarea mesajelor;

- *Reluarea* - se produce atunci când un mesaj sau o parte a acestuia este reluată (repetată), în intenția de a produce un efect neautorizat. De exemplu, este posibilă reutilizarea informației de autentificare a unui mesaj anterior;

- *Modificarea mesajelor* - face ca datele mesajului să fie alterate prin modificare, inserare sau ștergere. Poate fi folosită pentru a schimba beneficiarul unui credit în transferul electronic de fonduri sau pentru a modifica valoarea acelui credit. O altă utilizare poate fi modificarea câmpului destinatar/expeditor al poștei electronice;

- *Refuzul serviciului* - se produce când o entitate nu reușește să îndeplinească propria funcție sau când face acțiuni care împiedică o altă entitate de la îndeplinirea propriei funcții;

- *Repudierea serviciului* - se produce când o entitate refuză să recunoască un serviciu executat. Este evident că în aplicațiile de transfer electronic de fonduri este important să se evite repudierea serviciului atât de către emițător, cât și de către destinatar.

Având în vedere multitudinea posibilităților de atac asupra unui sistem de transmitere și stocare a datelor, o metodă de protejare a informațiilor, stocate sau transmise, ar putea fi criptarea.

# Introducere în criptografie

*Criptografie = κρυπτός {kryptós} (ascuns) + γράφειν {gráfein} (a scrie)*

Criptografia (cuvânt derivat din limba greacă a cuvintelor *kryptós* și *gráfein* reprezentând *scriere ascunsă*) este știința care se ocupă cu studiul *codurilor* și *cifrurilor*. Un cifru este de fapt un algoritm criptografic care poate fi folosit pentru a transforma un mesaj clar (*text clar*) într-un mesaj indescifrabil (*text cifrat*). Acest proces de transformare se numește *criptare* iar procesul invers se numește *decriptare*. Textul cifrat poate fi transmis ulterior prin orice canal de comunicații fără a ne face griji că informații sensibile ar putea ajunge în mâinile inamicilor.

Știința care se ocupă cu decriptarea (spargerea) cifrurilor se numește criptanaliză. Criptanaliza se ocupă cu studiul transformării unui text neinteligibil înapoi în cel inteligibil fără a cunoaște cheia de criptare.

Sistemul format dintr-un algoritm de criptare și o cheie de criptare se numește *criptosistem*.

Inițial, securitatea unui cifru depindea de faptul că inamicul nu cunoștea algoritmul de criptare folosit, dar pe măsură ce criptografia a evoluat, securitatea cifrului s-a bazat pe utilizarea unei chei secrete care se poate extrage din textul cifrat. Până la jumătatea secolului XX, nu a fost demonstrat faptul că un anumit cifru nu poate fi spart, ba chiar întreaga istorie a criptografiei este plină de relatări în care anumit cifru era spart iar ulterior erau creați alți algoritmi care la rândul lor erau sparți.

## Criptografia clasică

Toate criptosistemele pot fi impărțite în două tipuri: criptosisteme simetrice numite și clasice sau convenționale și criptosisteme asimetrice numite și moderne. Criptosistemele simetrice, sau cu cheie secretă, sunt acele criptosisteme în care numai emițătorul și receptorul cunosc cheia secretă pe care o aplică la fiecare criptare sau decriptare. Criptosistemele asimetrice, sau cu cheie publică, se bazează pe perechi de chei. Una din chei (cheia publică) este folosită la criptare, iar celaltă (cheia privată) este folosită la decriptare.

În criptografia clasică mesajul clar, numit și text clar, este convertit într-o secvență aparent aleatoare și fără sens, numită text cifrat. Procesul de criptare presupune un algoritm de criptare și o cheie de criptare. Această cheie este o valoare independentă de textul care se dorește a fi criptat. Odată produs, textul criptat trebuie transmis destinatarului. La recepție acest text criptat trebuie transformat în textul original folosind un algoritm de decriptare bazat pe aceeași cheie folosită la criptare.

### Securitatea criptării clasice

Securitatea criptării convenționale depinde de două aspecte esențiale: algoritmul de criptare și cheia de criptare. Algoritmul de criptare, care trebuie să fie destul de puternic pentru a face imposibilă o decriptare numai pe baza textului criptat. Cheia de criptare trebuie să fie destul de mare pentru a asigura o criptare puternică și mai ales trebuie să fie secretă. Deci nu este nevoie de păstrarea în secret a algoritmului folosit, ci numai a cheii de criptare. Acest lucru oferă un avantaj foarte mare criptării convenționale și a ajutat la răspândirea ei pe scară largă.

Există două cerințe esențiale care trebuie să le îndeplinească un algoritm de criptare :
1. Costul spargerii codului să depășească valoarea informației criptate;

2. Timpul necesar spargerii codului să depășească timpul de viață al informației, adică timpul până când informația are valoare.

Un algoritm de criptare care satisface aceste două cerințe este numit algoritm cu *securitate computațională*.

Prezentăm în tabelul 1, cât timp este necesar pentru a decripta un text cifrat, folosind metoda forței brute (brute force), pentru diferite dimensiuni ale cheii de criptare.

| Dimensiunea cheii (biți) | Numărul de chei posibile | Timpul necesar pentru 1 decriptare / µs | Timpul necesar pentru $10^6$ decriptări / µs |
|---|---|---|---|
| 64 | $2^{64} = 1.8 \times 10^{19}$ | $2^{63}$ µs = $2.9 \times 10^5$ ani | 106 zile |
| 128 | $2^{128} = 3.4 \times 10^{38}$ | $2^{127}$ µs = $5.4 \times 10^{24}$ ani | $5.4 \times 10^{18}$ ani |
| 256 | $2^{256} = 1.1 \times 10^{77}$ | $2^{255}$ µs = $1.8 \times 10^{63}$ ani | $1.8 \times 10^{57}$ ani |
| 512 | $2^{512} = 1.3 \times 10^{154}$ | $2^{511}$ µs = $6.7 \times 10^{153}$ ani | $6.7 \times 10^{147}$ ani |

Tabelul 1. Timpi necesari pentru decriptare folosind forța brută

Un algoritm de criptare este de *securitate necondiționată* dacă textul criptat generat de acesta nu conține destulă informație pentru a determina textul original, indiferent de volumul de text decriptat care este în posesia atacatorului. De asemenea, nu contează puterea de calcul și timpul de care dispune inamicul, mesajul nu poate fi decriptat deoarece informația necesară nu este acolo.

Cu excepţia unui algoritm numit "one-time pad", propus de Gilbert Vernam [2, 3] în 1920, nu există algoritm de criptare care să fie de o securitate necondiționată. Securitatea acestui algoritm a fost demonstrată în 1949 de către Claude Shannon, condițiile fiind ca, cheia de criptare să fie de aceeași lungime cu textul clar, să fie secretă și să nu fie folosită decât o singură dată [4]. Algoritmul „one-time pad" a fost implementat pentru transmiterea de informații sensibile, dar, marea problemă rămâne distribuirea cheilor de criptare care trebuiesc schimbate la fiecare utilizare și sunt foarte mari.

### Modelul criptării clasice

Un model de criptosistem simetric (clasic) este prezentat în figura 2.

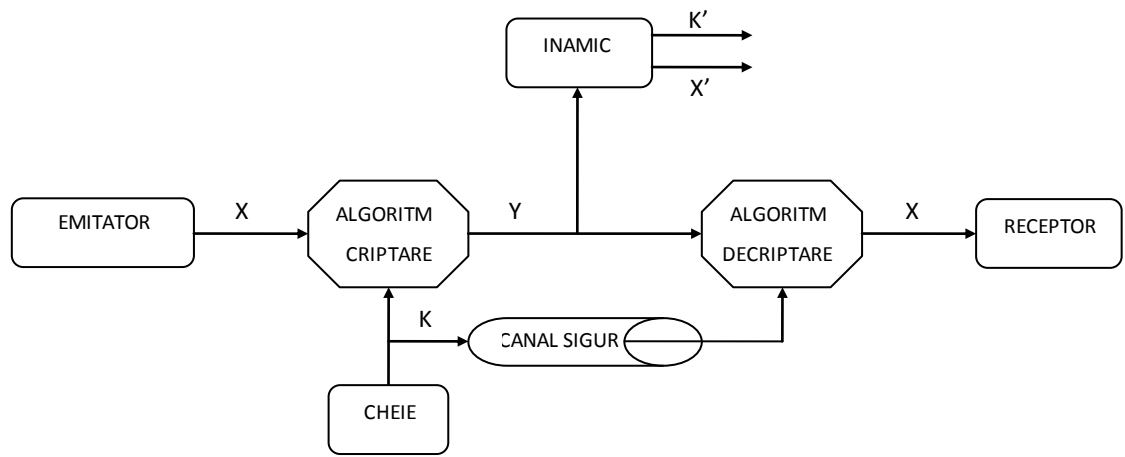

Figura 2. Un criptosistem clasic de comunicații

O sursă "emițător" produce un mesaj de tip text simplu (un șir de caractere), $X = [X1, X2, \ldots, X_M]$. Cele M elemente din X sunt niște litere ale unui alfabet finit. Pentru criptare trebuie generată o cheie de forma $K = [K_1, K_2, \ldots, K_J]$. Dacă cheia este generată de sursa mesajului aceasta trebuie transmisă destinatarului printr-un canal de comunicație sigur. O alternativă este ca o a treia parte să genereze cheia care să fie transmisă celor două părți folosind o metodă cu grad înalt de securitate.

Cu mesajul X și cheia de criptare K ca și intrare, algoritmul de criptare formează textul criptat $Y = [Y_1, Y_2, \ldots, Y_N]$. Astfel putem scrie:

$$Y = E_K(X)$$

Această notație indică faptul că Y este produs folosind algoritmul E pe baza textului X cu o funcție specifică determinată de valoarea cheii K.

Destinatarul, aflat în posesia cheii, poate inversa procesul de transformare:

$$X = D_K(Y)$$

unde D este algoritmul de decriptare.

Un inamic, observând Y și presupunând că cunoaște algoritmul de criptare E și decriptare D, va încerca să reconstituie X sau K sau ambele.

Procesul prin care se încearcă descoperirea lui X sau K, adică a textului necriptat, respectiv a cheii de criptare, se numește criptanaliză.

## *Criptografia modernă*

Un criptosistem asimetric care deocamdată este considerat a fi sigur poate fi implementat folosind algoritmul RSA [5], creat în 1978 de către Ronald Rivest, Adi Shamir și Leonard Adleman. Algoritmul RSA este folosit în prezent pentru securizarea comunicațiilor din internet, a tranzacțiilor bancare sau a comerțului electronic. Securitatea lui se bazează pe complexitatea matematică pe care o impune factorizarea numerelor prime.

Pentru securizarea comunicațiilor, guvernul Statelor Unite folosește algoritmul AES (Advanced Encryption Standard) [8], dezvoltat de către Joan Daemen și Vincent Rijmen și acceptat ca standard de către NIST (National Institute of Standards and Technology) în anul 2001. Algoritmul AES este un cifru bloc (128 biți) simetric capabil să cripteze sau să decripteze informația folosind chei criptografice pe 128,192, respectiv 256 de biți. AES se remarcă prin simplitate și prin performanțe criptografice ridicate, fiind ușor de implementat atât software cât și hardware.

Deci, criptosistemele cu chei publice suplinesc dezavantajul major al celor cu cheie secretă datorită faptului că nu mai este necesar schimbul de chei. Totuși, criptosistemele RSA și AES au marele inconvenient că securitatea lor se bazează pe complexitatea matematică a calculelor; în funcție de dimensiunea cheii folosite, pentru decriptare pot fi necesari și câteva mii de ani, la puterea de calcul actuală.

Având în vedere faptul că, încă din 1985, David Deutsch a descris principiile de funcționare ale unui calculator cuantic [9] – un supercalculator cu o putere de calcul extraordinar de mare care funcționează pe principiile fizicii cuantice, putem presupune că în viitor criptosistemele cu chei publice ar putea deveni nesigure.

În concluzie, singurul criptosistem absolut sigur rămâne one-time pad. Problema schimbului de chei poate fi rezolvată printr-un sistem de distribuire a cheilor cuantice (QKD – Quantum Key Distribuiton).

## Criptografia cuantică

Criptografia cuantică (Quantum Cryptography) este total diferită de criptografia convențională. Ea nu se bazează pe presupusa complexitate a unei probleme matematice, ci pe principiile fizicii cuantice – mai exact, pe *principiul incertitudinii al lui Heisenberg* [10]. Acest principiu spune că, dacă măsurăm o anumită proprietate cuantică, vom modifica într-o anumită măsură o altă proprietate cuantică.

### Principiul incertitudinii al lui Heisenberg :

Pentru oricare două observabile cuantice A și B avem :

$$\langle (\Delta A)^2 \rangle \langle (\Delta B)^2 \rangle \geq \frac{1}{4} \| \langle [A,B] \rangle \|^2,$$

unde

$$\Delta A = A - \langle A \rangle \quad \text{iar} \quad \Delta B = B - \langle B \rangle$$

și

$$[A, B] = AB - BA$$

Astfel, $\langle (\Delta A)^2 \rangle$ și $\langle (\Delta B)^2 \rangle$ sunt deviațiile standard care măsoară incertitudinea observabilelor A și B. Pentru observabilele A și B, pentru care $[A, B] \neq 0$, reducerea incertitudinii $\langle (\Delta A)^2 \rangle$ al observabilei A forțează creșterea incertitudinii $\langle (\Delta B)^2 \rangle$ al observabilei B și invers.

Folosind fenomene cuantice, se poate proiecta și implementa un sistem de comunicație care să evite întotdeauna interceptarea.

Având în vedere cele menționate, rezultă că, folosind principiile fizicii cuantice, putem realiza un sistem prin care se poate face un schimb de chei în deplină siguranță [1].

## Distribuirea cheilor cuantice

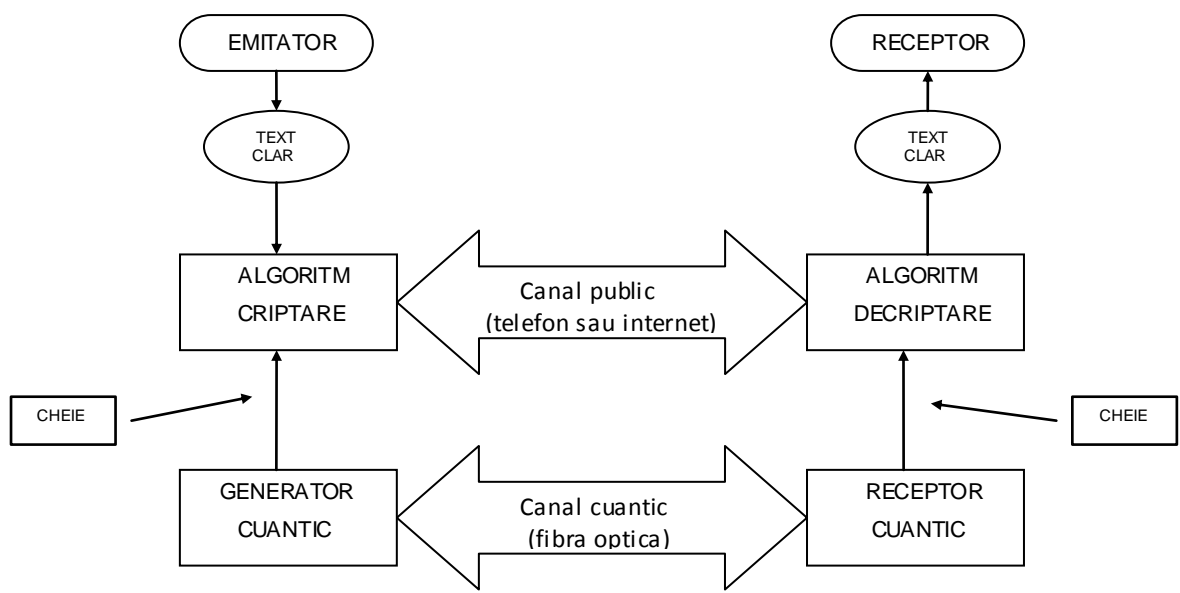

Figura 3. Un criptosistem cuantic de comunicații

Distribuirea cheilor cuantice folosește avantajul unor fenomene ce au loc la nivel subatomic, astfel încât, orice încercare de interceptare a biților, nu numai că eșuează dar și alertează *receptorul* că s-a produs o interceptare. În esență, fiecare bit din cheia transmisă, corespunde unei stări particulare a particulei purtătoare, cum ar fi fotonii polarizați. *Emițătorul* cheii trebuie să stabilească o secvență de polarizare a fotonilor, care vor fi transmiși printr-o fibră optică. Pentru a obține cheia, care este formată dintr-o secvență de fotoni polarizați - qbits, *Receptorul* trebuie să facă o serie de măsurători cu ajutorul unor filtre cu care se poate determina polarizarea fotonilor.

Un foton poate fi polarizat *liniar* ($0°$, $90°$), *diagonal* ($45°$, $135°$) sau *circular* (stânga - spinL, dreapta - spinR). Polarizarea liniară-diagonală, liniară-circulară sau diagonală-circulară sunt cunoscute ca fiind *variabile legate* iar legile fizicii cuantice spun că este imposibil de măsurat simultan valorile oricărei perechi de *variabile legate;* deci, dacă *inamicul* încearcă să „măsoare" un foton polarizat orizontal, folosind o metoda de „măsurare" a fotonilor polarizați diagonal, acel foton își schimbă polarizarea din orizontală în diagonală.

| simbolizare | L(Liniar) | D(Diagonal) | S(Stanga) | R(Right) | L(Liniar) | D(Diagonal) |
|---|---|---|---|---|---|---|
| polarizare | 0° | 45° | spinL | spinR | 90° | 135° |
| qbit | 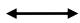 | 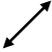 | 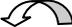 | 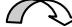 | 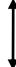 | 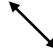 |
| bit | 0 | 0 | 0 | 1 | 1 | 1 |

Tabelul 2. Polarizarea fotonilor - qbits

## Protocolul BB84 – QKD (Quantum Key Distribution)

Protocolul BB84, este primul protocol cuantic, numit astfel după cei care l-au descoperit în 1984, Charles Bennett and Gilles Brassard, prin care două entități, *emițătorul* și *receptorul* stabilesc în secret o cheie unică, comună, folosind fotoni polarizați (*qbits*).

*Protocolul BB84 :*

1. *Emițătorul* generează o secvență aleatoare de biți - notată "*s*".
2. *Emițătorul* alege aleator ce bază de polarizare să aplice fiecărui foton din "*s*" (liniar "R" sau diagonal "D"). Notăm "b" secvența de polarizare.
3. *Emițătorul*, folosind un echipament special (un laser atenuat și un set de dispozitive de polarizare a fotonilor), crează o secvență "p" de fotoni polarizați – qbits a căror direcție de polarizare reprezintă biții din "*s*".
4. *Emițătorul* transmite qbits din "p" prin fibră optică către *Receptor*.
5. *Receptorul*, pentru fiecare qbit recepționat, alege aleator câte o bază de polarizare (liniar "R" sau diagonal "D"). Notăm cu "b'" secvența bazelor de polarizare aleasă.
6. *Receptorul*, măsoară fiecare qbit recepționat respectând baza de polarizare aleasă la pct. 5, rezultând o secvență de biți "*s'*".
7. *Emițătorul* îi transmite printr-un canal public *Receptorului*, ce bază de polarizare a ales pentru fiecare bit în parte. La rândul sau *Receptorul* îi comunică *Emițătorului* unde a făcut aceeași alegere a bazei de polarizare. Biții pentru care cei doi nu au avut aceeași bază de polarizare sunt eliminați din "*s*" și "*s'*".

*Detectarea inamicului :*

Pentru bitul cu numărul *n*, bitului *s*[n] îi va corespunde o bază de polarizare, b[n] iar bitului *s'*[n] îi va corespunde o bază b'[n] .

Dacă   b'[n] = b[n]    va implica că    *s'*[n] = *s*[n].

În cazul în care un inamic a încercat să citească fotonul purtător al lui *s*[n], atunci chiar dacă cele două baze alese de *Receptor* și *Emițător* sunt identice (b'[n] = b[n]), vom avea *s'*[n] ≠ *s*[n]. Aceasta le va permite *Emițătorului* și *Receptorului* să detecteze prezența *Inamicului* și să reprogrameze transmisia.

*Secret key reconciliation :*

Algoritmul BB84 este incomplet datorită faptului că, chiar dacă *Inamicul* este prezent sau nu, vor exista erori în secvența de biți a *Receptorului*.

Pasul final al algoritmului BB84 constă într-o comparație dintre cele două secvențe de biți, deținute de *Emițător* și *Receptor* după codificare și decodificare. Acesta cuprinde două etape : *secret key reconciliation* [6] și *privacy amplification* [6, 7].

*Etapa secret key reconciliation este o procedură de corectare a erorilor care elimină :*

- *erorile generate de alegerea diferită a bazelor*
- *erorile generate de Inamic*
- *erorile generate de zgomote*

*Etapa Secret key reconciliation realizează o căutare binară, interactivă a erorilor. Emițătorul* și *Receptorul* împart secvența de biți rămasă în blocuri de biți și vor compara paritatea fiecărui bloc. În cazul în care paritatea unui bloc de biți diferă, ei vor împărți blocul respectiv în blocuri mai mici și vor compara paritatea lor. Acest proces va fi repetat până când bitul care diferă va fi descoperit și eliminat. Comunicațiile din această etapă se vor realiza pe un canal public nesecurizat.

*Privacy amplification :*

Având în vedere faptul că, în etapa anterioară, comunicația s-a realizat pe un canal nesecurizat, există posibilitatea ca *Inamicul* să dețină informații sensibile despre cheia secretă. Pentru a stabili o cheie perfect sigură, *Emițătorul* și *Receptorul* trebuie să mai realizeze o etapă : *privacy amplification. Această etapă constă într-o permutare a biților din cheia secretă și eliminarea unui subset de biți, care va fi realizată de către Emițător și Receptor.*

La finalul acestei etape avem certitudinea că Emițătorul și Receptorul, dețin aceeași cheie secretă care nu este cunoscută de Inamic.

2009 - Anghel CătălinPagina **20** din **24**

## Concluzii și direcții de cercetare

În concluzie, algoritmul BB84, ne oferă posibilitatea stabilirii unei chei unice și sigure între un emițător și un receptor. Această cheie poate fi folosită pentru a putea realiza o comunicație securizată împreună cu algoritmul de criptare „one-time pad". Totuși, algoritmul de criptare „one-time pad" coroborat cu algoritmul de distribuire a cheilor cuantice BB84 are a mare slăbiciune, în sensul că securitatea lui se bazează pe complexitatea matematică a calculelor necesare pentru decriptarea lui. Este cunoscut faptul că, pe plan mondial, ca aplicații ale fizicii cuantice, oamenii de știință lucrează la realizarea unui calculator cuantic, calculator ce va folosi fotonii în locul electronilor. Se estimează că acest calculator, atunci când va fi realizat, va putea decripta în timp real toate comunicațiile criptate transmise prin internet. Plecând de la această premisă, se impune realizarea unui algoritm de criptare și transmisie a datelor, a cărui siguranță să nu se bazeze pe complexitatea calculelor matematice ci pe principiile fizicii cuantice.

## Raportarea rezultatelor cercetării

Pe parcursul derulării programului de cercetare științifică, se vor elabora și prezenta public trei rapoarte de cercetare, după cum urmează :

- ➢ Raportul 1 : Stadiul actual și direcții de cercetare în criptografia cuantică.
    - ✓ Obiective :
        - Prezentarea stadiului actual de dezvoltare în criptografia cuantică;
        - Prezentarea direcțiilor de cercetare;
    - ✓ Activități :
        - Documentare;
        - Diseminare;
- ➢ Raportul 2 : Contribuții aduse criptografiei cuantice prin tehnici și protocoale de selecție a bazelor și sincronizare a transmisiei.
    - ✓ Obiective :
        - Realizarea unui algoritm îmbunătățit de distribuție a cheilor cuantice;
    - ✓ Activități :
        - Documentare;
        - Implementarea algoritmului;
        - Programarea algoritmului în C++;
        - Diseminare;
- ➢ Raportul 3 : Contribuții aduse criptografiei cuantice prin tehnici și metode pentru detecția atacurilor în transmisiile de date.
    - ✓ Obiective :
        - Realizarea unui algoritm îmbunătățit de detectare a inamicului;
    - ✓ Activități :
        - Documentare;
        - Implementarea algoritmului;
        - Programarea algoritmului în C++;
        - Diseminare;

Calendarul estimat al prezentării rapoartelor de cercetare este urmatorul:

    Raportul 1 va fi prezentat în februarie 2010

    Raportul 2 va fi prezentat în septembrie 2010

    Raportul 3 va fi prezentat în septembrie 2011

Diseminarea rezultatelor cercetării se va face prin participarea la conferințe naționale sau internaționale de specialitate și prin publicarea unor articole în reviste de profil cu un grad de difuzare cât mai mare. O parte din rezultate vor fi prezentate ca lucrări de laborator pentru studenți, în activitatea didactică în cadrul catedrei.

Activitatea de diseminare a rezultatelor a demarat deja prin publicarea a două lucrări la conferințe internaționale [11], [12] și va continua pe tot parcursul cercetării doctorale.

    **Titlul tezei**

    Contribuții aduse la

    **Securizarea Sistemelor Informatice și de Comunicații**

    prin

    **Criptografia Cuantică**

## Referințe

[1]. Bennett C. H. & Brassard G. (1984). "Quantum cryptography: Public key distribution and coin tossing". Proceedings of IEEE International Conference on Computers Systems and Signal Processing, Bangalore India, December 1984, pag. 175-179.

[2]. G. Vernam, "Secret signaling system," U.S. patent No. 1310719 (applied in 1918, granted in 1919).

[3]. G. Vernam, "Cipher printing telegraph system for secret wire and radio telegraphic communications," J. Am. Institute of Electrical Engineers Vol. XLV, pag. 109–115 (1926).

[4]. C. Shannon, "Communication theory of secrecy systems," Bell System Technical J. 28, pag. 656–715 (1949).

[5]. R.L. Rivest, A. Shamir, and L. Adleman, "A method for obtaining digital signatures and public-key cryptosystems," Comm. ACM 21, pag. 120–126 (1978).

[6]. Bennett C.H., Bessette, F., Brassard, G., Salvail, L., & Smolin, J., "Experimental quantum cryptography", J. Cryptology, vol. 5, no. 1, 1992, pag. 3 – 28.

[7]. Bennett Charles H., Gilles Brassards, and Jean-Marc Roberts, "Privacy amplification by public discussions", Siam J. Comput, Vol 17, No. 2, April 1988, pag. 210 -229.

[8]. Joan Daemen and Vincent Rijmen, "The Design of Rijndael: AES - The Advanced Encryption Standard." Springer-Verlag, 2002.

[9]. Deutsch David (July 1985). "Quantum theory, the Church-Turing principle and the universal quantum computer". Proceedings of the Royal Society of London; Series A, Mathematical and Physical Sciences 400 (1818): pag. 97–117.

[10]. Heisenberg Werner (1927), "About the perceptual content of quantum kinematics and mechanics ", Journal of Physics 43, pag: 172–198.

## Articole și lucrări de diseminare a cercetării